\documentclass[a4paper]{ifacconf}

\usepackage{graphicx,amsmath,url}      
\usepackage[round]{natbib}   

\def\portugues{1} 

\ifx\portugues\undefined
\def\portugues{0}
\fi

\if\portugues0
   \usepackage[english]{babel}
  \else
   \usepackage[spanish,brazil,english]{babel}
\fi

\usepackage[T1]{fontenc}

\usepackage[utf8]{inputenc}
\usepackage{gensymb}
\usepackage{ae}

\if\portugues1
 { }
 { }
 { }
 
\fi

\begin{document}

\if\portugues1

%
\selectlanguage{brazil}
	
\begin{frontmatter}

\title{Detecção de Fadiga Muscular Utilizando Modulação Variável de Fluxo de Bits e Correlação Cruzada de Sinais Eletromiográficos} 



\author[First]{Leandro C. Medeiros} 
\author[Fourth]{Pedro H. O. Silva} 
\author[Second]{Victor H. S. Lopes}
\author[Fourth]{Arthur F. Oliveira} 
\author[Second]{Eduardo B. Pereira}
\author[First]{Erivelton G. Nepomuceno}

\address[First]{Programa de Pós-Graduação em Engenharia Elétrica (UFSJ/CEFET-MG), GCOM - Grupo de Controle e Modelagem,  Universidade Federal de São João del-Rei, MG, \\ (e-mail: leandro.medeiros@mrs.com.br, nepomuceno@ufsj.edu.br).}
\address[Second]{Núcleo de Tecnologias Assistivas (CyRoS), São João del-Rei, MG, (e-mail: victorldw@gmail.com, ebento@ufsj.edu.br).}
\address[Fourth]{ Programa de Pós-Graduação em Engenharia Elétrica (UFJF-MG), Universidade Federal de Juiz de Fora, MG,\\ (e-mail: pedrolives@hotmail.com.br, arthur.filgueiras@mrs.com.br).}

\selectlanguage{english}
\renewcommand{\abstractname}{{\bf Abstract:~}}
\begin{abstract}                
The surface electromyography (sEMG) analysis can provide information on muscle fatigue status by estimation of muscle fibre conduction velocity (MFCV), a measure of the travelling speed of motor unit action potentials in muscle tissue. This paper proposes a technique for MFCV estimation using cross-correlation methods and variable bitstream modulation. The technique displays an estimate based on a set of data generated by the gain variation modulation, providing an average estimate of the MFCV. The observed trend of MFCV decrease correlates with the fatigue state of the observed muscle. Finally, the values found were compared with information from the literature, validating the method and showing the advantages of using variable modulation.

\vskip 1mm
\selectlanguage{brazil}
{\noindent \bf Resumo}: A análise da eletromiografia de superfície (EMGs) pode fornecer informações sobre o estado de fadiga muscular pela estimação da velocidade de condução de fibras musculares ($\mathrm{VCFM}$), sendo uma medida da velocidade de deslocamento dos potenciais de ação no tecido muscular. O presente artigo propõe uma técnica para estimativa da $\mathrm{VCFM}$ utilizando métodos de correlação cruzada e modulação variável de fluxo de bits. A técnica exibe uma estimativa baseada em um conjunto de dados gerados pela variação do ganho de modulação, fornecendo uma estimativa média da $\mathrm{VCFM}$. A estimativa  representa uma diminuição da $\mathrm{VCFM}$ correlacionada com o estado de fadiga do músculo observado. Por fim, os valores encontrados foram comparados com informações da literatura, validando o método e mostrando as vantagens do uso da modulação variável.

\end{abstract}

\selectlanguage{english}


\selectlanguage{brazil}

\end{frontmatter}
\else
%

\begin{frontmatter}

\title{Style for SBA Conferences \& Symposia: Use Title Case for
  Paper Title\thanksref{footnoteinfo}} 

\thanks[footnoteinfo]{Sponsor and financial support acknowledgment
goes here. Paper titles should be written in uppercase and lowercase
letters, not all uppercase.}

\author[First]{First A. Author} 
\author[Second]{Second B. Author, Jr.} 
\author[Third]{Third C. Author}

\address[First]{Faculdade de Engenharia Elétrica, Universidade do Triângulo, MG, (e-mail: autor1@faceg@univt.br).}
\address[Second]{Faculdade de Engenharia de Controle \& Automação, Universidade do Futuro, RJ (e-mail: autor2@feca.unifutu.rj)}
\address[Third]{Electrical Engineering Department, 
   Seoul National University, Seoul, Korea, (e-mail: author3@snu.ac.kr)}
   
\renewcommand{\abstractname}{{\bf Abstract:~}}   
   
\begin{abstract}                
These instructions give you guidelines for preparing papers for IFAC
technical meetings. Please use this document as a template to prepare
your manuscript. For submission guidelines, follow instructions on
paper submission system as well as the event website.
\end{abstract}


\end{frontmatter}
\fi

\section{Introdução}

A fadiga, em sentido amplo, ocorre na vida cotidiana e pode ser descrita como uma sensação de fraqueza, dor muscular ou uma diminuição no desempenho durante atividades físicas ou cognitivas \citep{Marco2017}. Na análise do sistema neuromuscular, a fadiga muscular é a redução gradual ou transitória na capacidade de geração de força pelos músculos. Durante as contrações submáximas sustentadas ou repetitivas, ocorrem alterações periféricas e centrais na atividade dos músculos submetidos à fadiga. Essas alterações são denominadas manifestações mioelétricas de fadiga, podendo ser avaliadas de forma não-invasiva mediante eletromiografia de superfície (EMGs). 

O sinal EMGs é de origem biológica e mede as correntes elétricas geradas nos músculos durante uma contração, representando as atividades neuromusculares geridas pelo sistema nervoso \citep{DeLuca1997,Loterio2017}. Os músculos são uma coleção fibras musculares, em que as unidades motoras consistem em neurônios motores juntamente com as fibras musculares inervadas. Os neurônios motores podem estimular as fibras musculares, ocasionando um aumento no campo elétrico ao redor da fibra. A estimulação ocorre devido ao movimento de \textit{íons} de potássio ($K^{+}$) e \textit{íons} de sódio ($Na^{+}$) nas células \citep{Hall2010}. A mudança no potencial elétrico provoca a contração e sua ausência o relaxamento dos músculos, em que esses processos eletroquímicos de estimulação originam potenciais de ação da unidade motora (MUAPs) \citep{Ergeneci2018}.  

O sinal detectado na superfície da pele é resultado da sobreposição de diferentes MUAPs que ocorrem durante as contrações dos músculos. Além disso, a velocidade de propagação do potencial de ação é chamada de velocidade de condução de fibra muscular ($\mathrm{VCFM}$). A velocidade de condução ao longo das fibras musculares é um parâmetro fisiológico relacionado ao tipo das fibras musculares, temperatura muscular, concentração de \textit{íons} e taxa de disparo das unidades motoras \citep{Xu2017}. Durante um processo de fadiga muscular, a $\mathrm{VCFM}$ exibe valores altos que reduzem progressivamente para valores baixos. Desta forma, a tendência observada da diminuição da $\mathrm{VCFM}$ correlaciona com o estado de fadiga do músculo observado. A diminuição  da $\mathrm{VCFM}$ ocorre devido ao acúmulo de \textit{íons} $K^{+}$ e ácido lático no espaço muscular extracelular, inibindo a condução dos potenciais de ação ao longo da membrana muscular, diminuindo a velocidade dos MUAPs.

Diversos métodos foram propostos na literatura para a estimativa da $\mathrm{VCFM}$ mediante sinais de EMGs. Entre os métodos são apresentadas técnicas que verificam os valores da frequência média do músculo, fragmentando o sinal no tempo para averiguação do decaimento da frequência média, indicando a fadiga local. Por exemplo, as técnicas que utilizam Transformada de Fourier e Transformada Wavelet Contínua para descrever as mudanças temporais no espectro de potência dos sinais EMGs \citep{Beck2005,Oliveira2013}. Outra abordagem para estimar a $\mathrm{VCFM}$ envolve o processamento de sinais EMGs derivados de dois eletrodos fixados ao longo da fibra muscular. Esses métodos assumem que os dois sinais detectados são análogos com incorporação de atraso. Sendo que a correlação cruzada dos sinais pode ser usada como estimador de atraso, em que incluindo a distância de separação dos eletrodos é possível o cálculo da velocidade de condução. 

Na literatura, existem métodos que utilizam a correlação cruzada aplicada em diferentes sistemas de detecção. Cada análise proposta possui a utilização de um método em combinação com um sistema de detecção específico, fornecendo uma definição de atraso e diferentes resultados da $\mathrm{VCFM}$ \citep{Farina2004}. Desta forma, a utilização do método de correlação cruzada deve ser associado a outro tipo de análise, que incorpora uma indicação mais conservadora. Uma possibilidade é considerar que os métodos possuem um grau de incerteza, principalmente derivados das diferentes sensibilidades ao ruído na aquisição dos dados. A incerteza pode ser incorporada considerando que o grau de correlação entre os sinais é variável, devido que ainda existem componentes de ruídos nos sinais.


Neste trabalho é proposto a utilização do método de correlação cruzada integrado à modulação variável de fluxo de bits, incorporando incerteza nos resultados. Desta forma, é investigado se a modulação do sinal melhora a estimativa da velocidade de condução do sinal EMGs. Uma vez que a estimativa do atraso depende fortemente do grau de similaridade dos sinais, a modulação possibilita obter representações da EMGs com faixas de valores de correlação ou similaridade. O conjunto estimado de velocidades possibilita o cálculo de médias, fornecendo estatisticamente uma estimativa que considera incertezas da velocidade a cada segundo. Foram obtidos resultados que comprovam a diminuição da velocidade de condução em um processo de fadiga, em que os dados foram comparados com valores fornecidos pela literatura que possuem protocolos experimentais semelhantes.

\section{Conceitos Preliminares}

\subsection{Correlação Cruzada}

Em processamento de sinais, a relação cruzada ou correlação cruzada é uma medida de similaridade entre dois sinais em função de um atraso \citep{Smith1999}. O método de correlação cruzada é comumente utilizado no reconhecimento de sinais incorporados em outros sinais. A correlação cruzada determina o grau de similaridade, como também a diferença de fase ou atraso entre dois sinais. O método é utilizado em aplicações de reconhecimento de padrões, rede neurais \citep{Ma2016}, criptoanálise e engenharia biomédica \citep{Sun2016}. A equação \ref{eq:correlação} define a correlação cruzada de tempo discreto:
\begin{equation}
\label{eq:correlação}
    r_{xy}[l]=\sum_{k=1}^{n}x[k]y[k-l],
\end{equation}
\noindent em que $l$ é o deslocamento de tempo entre os dois sinais (tempo de atraso discreto), $k$ é a amostra avaliada e $n$ é o comprimento da janela de correlação, na qual todas as amostras são multiplicadas e progressivamente somadas.

\subsection{Modulação}

A obtenção de um sinal modulado em fluxo de bits envolve a aplicação de quantização. Na quantização, uma amostra com uma determinada amplitude é convertida em um novo conjunto de valores quantizados. A modulação é feita dividindo um intervalo de valores em níveis diferentes e atribuindo uma amplitude a cada nível no intervalo de quantização. O termo LSB defini o passo de quantização e representa a distância entre os níveis de quantização, representado por:
\begin{equation}
    \mathrm{LSB}=\frac{V_{\mathrm{max}}}{2^N},
\end{equation}
\noindent em que $V_{\mathrm{max}}$ é a excursão máxima do sinal e $N$ é o número de \textit{bits} usados na quantização.

\section{Metodologia}


A modulação em fluxo de bits possibilita condicionar os sinais de EMGs mediante a variação do nível de quantização, obtendo variações da sua representação. Desta forma, as diferentes representações dos sinais em fluxo de bits resultam em atrasos distintos, caracterizando um intervalo de velocidades de condução dos dois sinais. O sinais de EMGs são processados utilizando o software Matlab R2016a\textsuperscript{\textregistered} e um computador com as seguintes configurações: processador Intel(R) Core(TM) i5-7300HQ CPU @ 2.5GHz - RAM 8 GB - Windows 10 Home Single Language.  

\begin{figure}[!ht]
\begin{center}
\includegraphics[width=7.2cm]{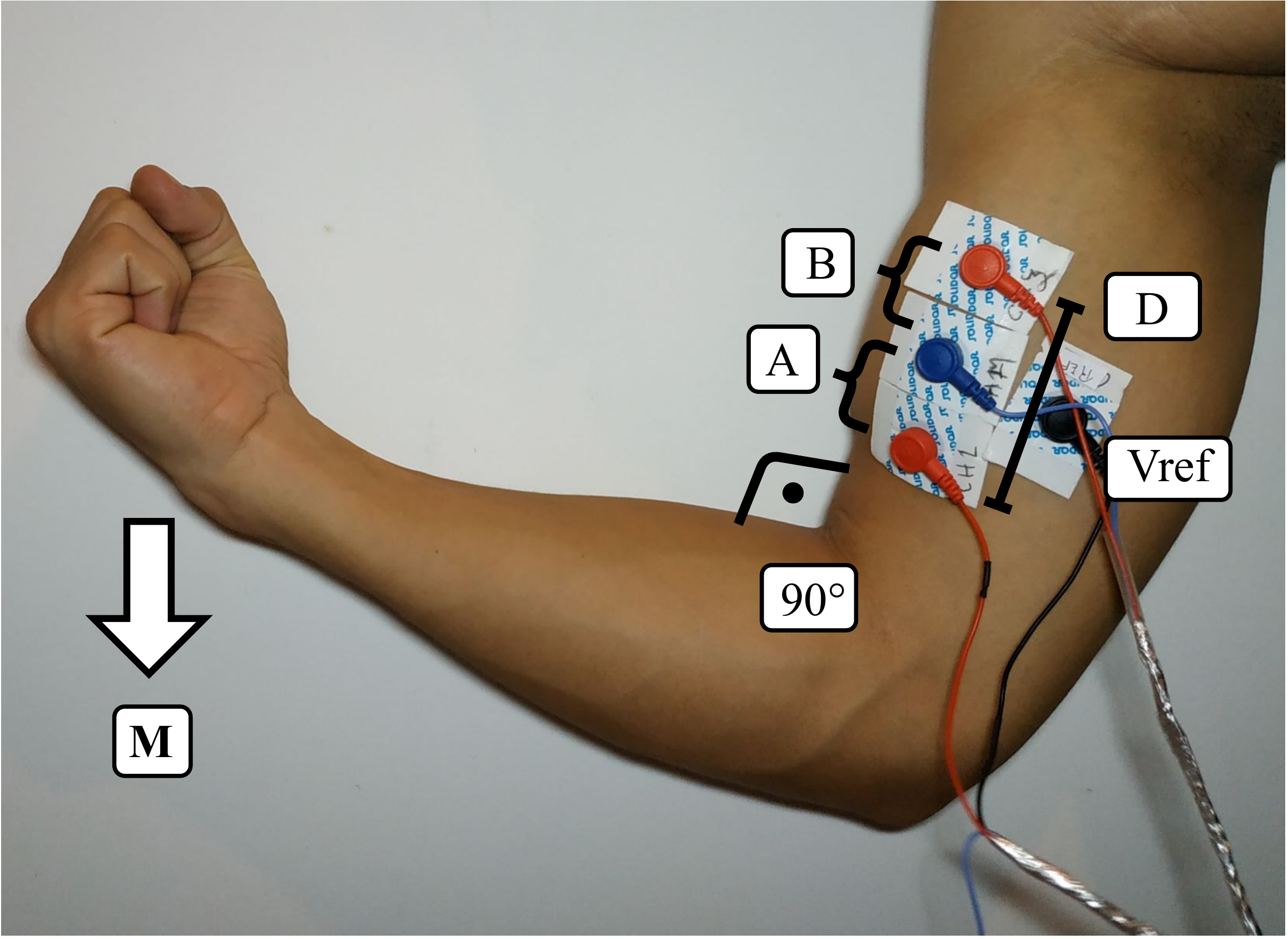}    
\caption{Ilustração do protocolo experimental, mostrando os dois canais (eletrodos) igualmente espaçados por uma distância D e fixados no bíceps braquial direito, também representando o ângulo de articulação do cotovelo e o movimento de rosca direta.} 
\label{fig:ilustra_exp}
\end{center}
\end{figure}

\subsection{Aquisição de Dados EMGs}

\begin{figure}[!ht]
\begin{center}
\includegraphics[width=8cm]{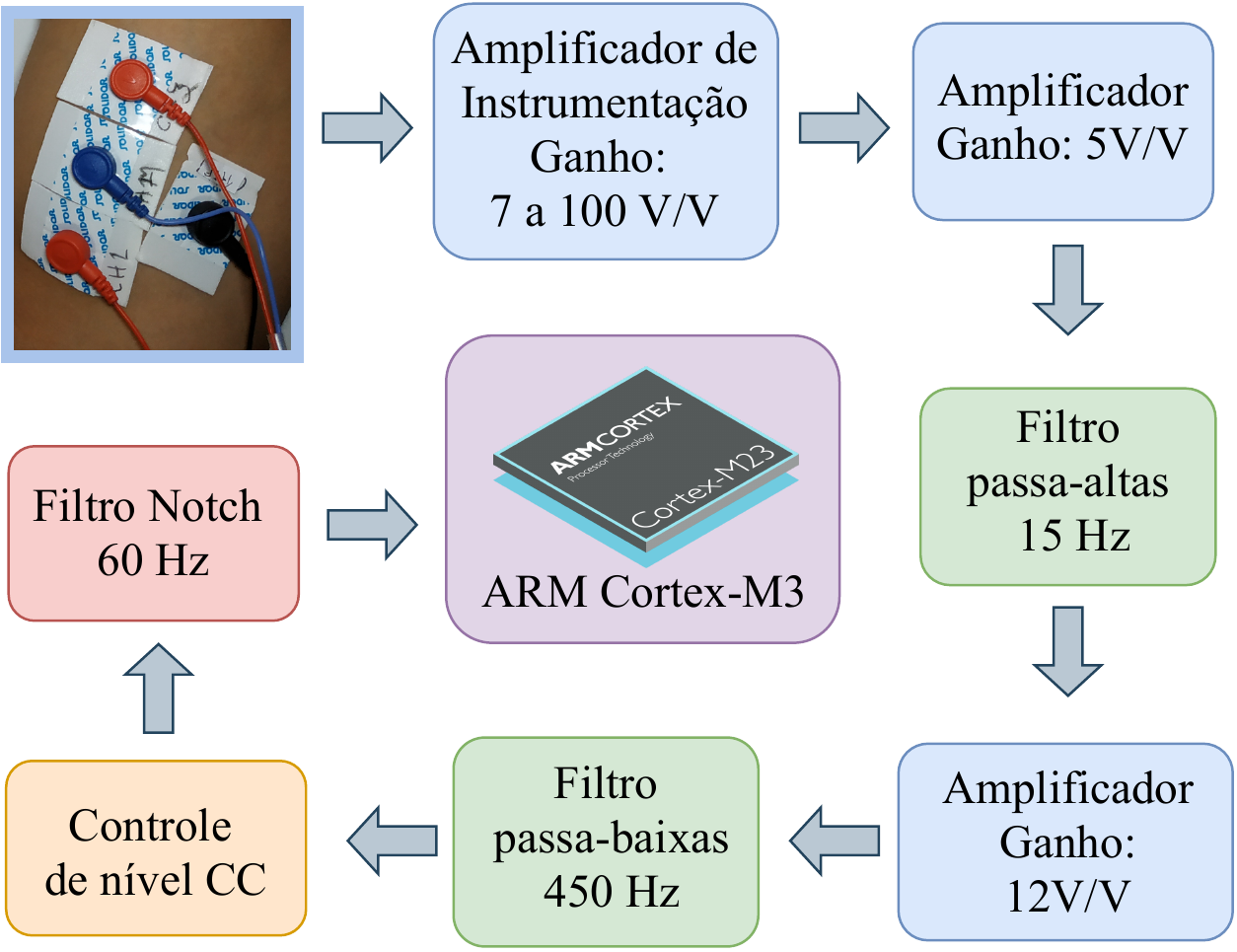}    
\caption{Diagrama esquemático do sistema de aquisição de EMGs, representando os estágios de condicionamento do sinal, começando dos eletrodos e finalizando no microcontrolador.}
\label{fig:diag_aquisição}
\end{center}
\end{figure}

Para a aquisição dos sinais foi utilizado um sistema de aquisição de sinais de EMGs. O sistema é composto por elementos de aquisição e condicionamento dos sinais para remover artefatos e ruídos da rede \citep{Lopes2018}. A Figura \ref{fig:diag_aquisição} mostra os estágios de amplificação e filtragem, a começar dos eletrodos e finalizando na aquisição do sinal condicionado realizada pelo microcontrolador ARM Cortex-M3\textsuperscript{\textregistered}. A coleta dos sinais EMGs foi realizada utilizando um ADC de $12$ \textit{bits} com uma frequência de amostragem de $2200$ $\mathrm{Hz}$ \citep{Lopes2017}.

\subsection{Protocolo Experimental}

    O objetivo dos experimentos é induzir o processo de fadiga muscular para avaliação do método proposto. Para induzir a fadiga foi escolhido inicialmente os músculos flexores do cotovelo, especificamente o bíceps braquial direito. O exercício escolhido para analisar o músculo foi o rosca direta, pois envolve principalmente o bíceps braquial e também consiste em um movimento limitado em termos de amplitude e de planos de movimento. Desta forma, foram escolhidos alguns parâmetros para analisar o processo de fadiga, que consistem no tipo de contração, ângulo da articulação do cotovelo (isométrico) e peso do \textit{halter} posicionado na mão (veja Figura \ref{fig:ilustra_exp}).
    
    A medição de eletromiografia foi realizada por quatro eletrodos colocados na pele do braço do participante, fixados na direção da fibra muscular. O braço foi posicionado em um apoio afim de minimizar movimentos compensatórios, realizando contrações segurando um halter por $34$ segundos. A Tabela \ref{tab:param_exp} representa os dados do protocolo de coleta utilizado, incluindo o tempo médio de descanso entre os experimentos.

\subsection{Modulação em Fluxo de Bits}

A modulação em fluxo de bits permite caracterizar os sinais de EMGs em valores discretos com níveis de quantização variáveis. O objetivo é usar a técnica para estabelecer diferentes condicionamentos para os dois sinais EMGs, obtendo intervalos da velocidade de condução. Desta forma, os intervalos auxiliam na avaliação da progressão de diminuição da velocidade de condução. Assim, a tendência observada da diminuição da $\mathrm{VCFM}$ correlaciona com o estado de fadiga do músculo ativo.

\begin{figure}[ht]
	\begin{minipage}[c][0.5\width]{
	   0.5\textwidth}
	   \centering
	   \includegraphics[width=1\textwidth]{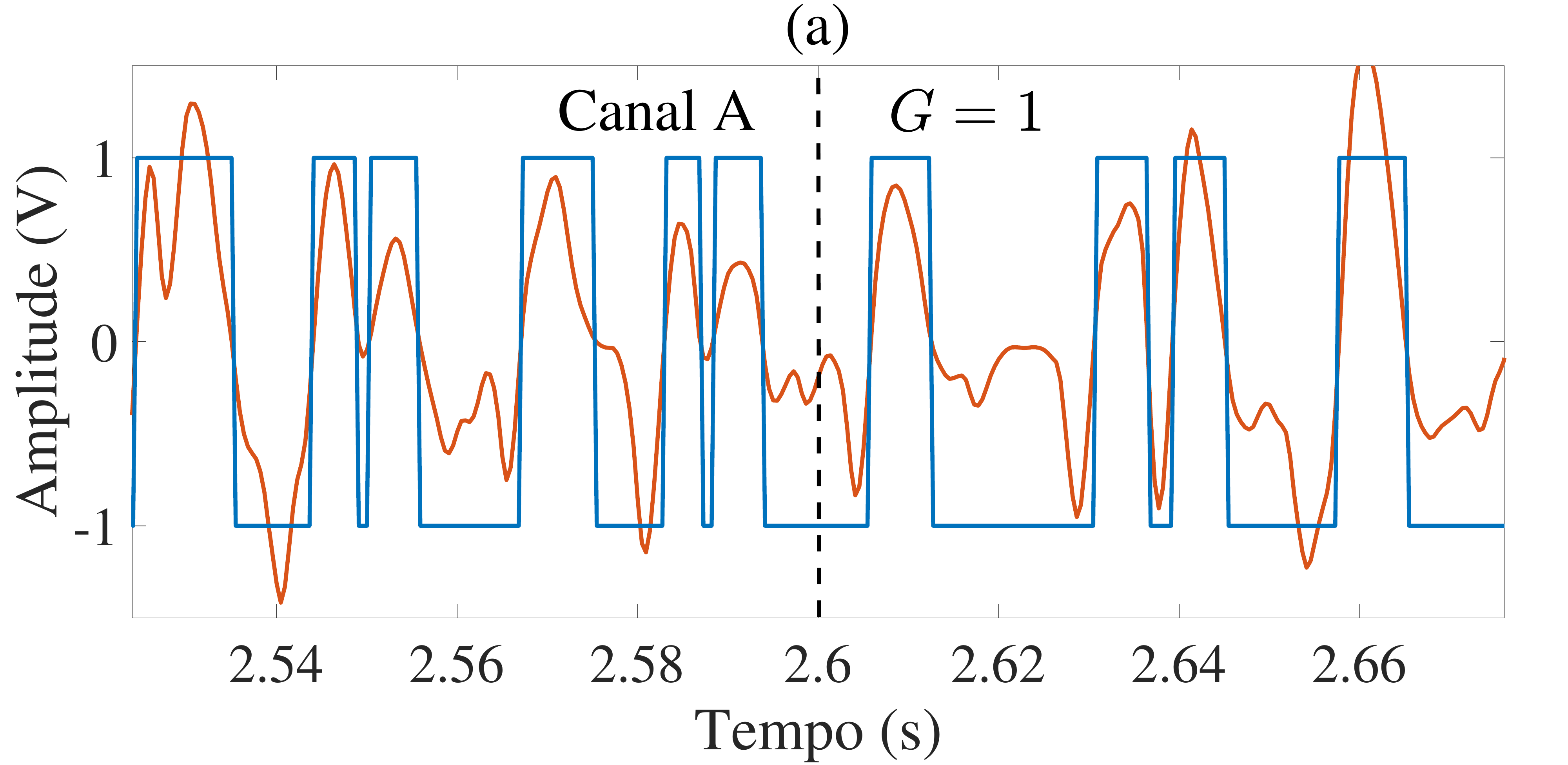}
	\end{minipage}
	\begin{minipage}[c][0.5\width]{
	   0.5\textwidth}
	   \centering
\includegraphics[width=1\textwidth]{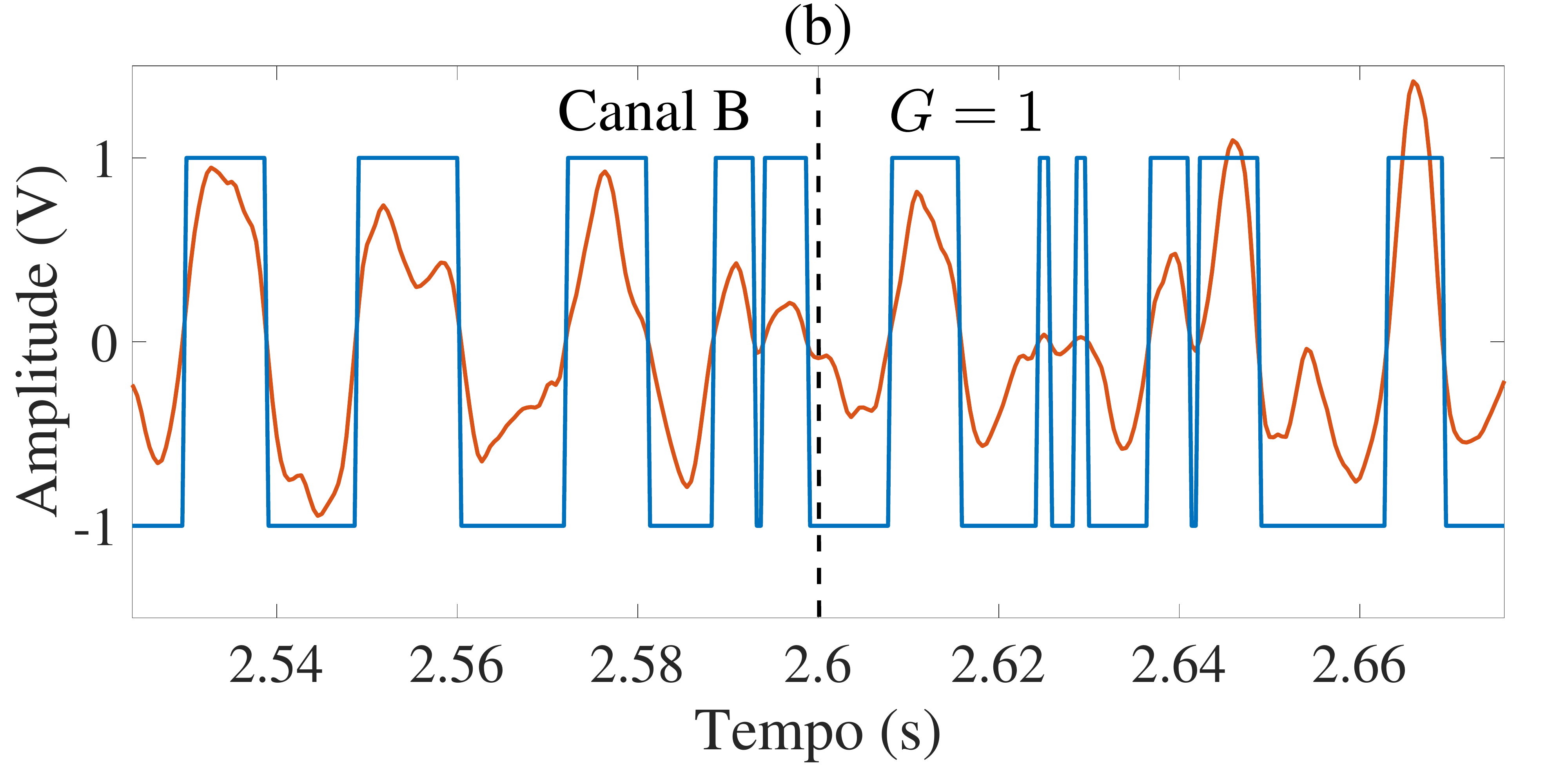}
	\end{minipage}
\caption{Modulação em fluxo de bits dos sinais EMGs coletados, representando o atraso entre os sinais no tempo de $2,6$ segundos como referência, (a) representa o canal A com ganho de modulação $G=1$ e (b) representa o canal B com ganho de modulação $G=1$.}
\label{fig:mod_comp_ab}
\end{figure}

Os sinais EMGs são amostrados e armazenados no Matlab\textsuperscript{\textregistered} e posteriormente são processados por um algoritmo de modulação em fluxo de bits com intervalo de quantização variável. O algoritmo define os intervalos de quantização relacionados aos valores da amplitude do sinal EMGs, sendo que $1$ corresponde a uma amplitude positiva e $-1$ corresponde a uma amplitude negativa (veja Figura \ref{fig:mod_comp_ab}). Um vetor é definido de tal forma que uma série de comparações é realizada para verificar os bits fixados para cada amostra, comportando como um detector de picos de EMGs.

A variação do intervalo de quantização configura diferentes fluxos de bits para uma mesma forma de onda de EMGs, em que a variação do intervalo é representada por um ganho de modulação ($G$). A mudança no valor do ganho torna o sistema mais receptivo, ou seja, mais picos podem ser detectados com o aumento do ganho, no entanto o sistema se torna mais propenso a captar ruídos contidos nos sinais.  A Figura \ref{fig:mod_comp_ab} representa os sinais modulados em fluxo de bits utilizando $G=1$ e comprova à existência de atraso entre os sinais derivados dos canais A e B pela referência no tempo de 2,6 segundos.

\begin{table}[!ht]
\centering
\setlength{\tabcolsep}{9pt} 
\renewcommand{\arraystretch}{1.1}
\caption{Parâmetros do protocolo realizado para coleta dos dados.}
\label{tab:param_exp}
\begin{tabular}{lccc}
\hline
\textbf{Parâmetros }                                                               & \multicolumn{3}{c}{\textbf{Valores}}                 \\ [1pt] \hline
$\mathrm{N}^{\circ}$ de Canais                                                                    & \multicolumn{3}{c}{$2$}                       \\ [1pt] \hline
\begin{tabular}[c]{@{}l@{}}Distância dos \\ canais\end{tabular}                                                                            & \multicolumn{3}{c}{$2,5$ $\mathrm{cm}$}                       \\ [1pt] \hline
\begin{tabular}[c]{@{}l@{}}Tipo de \\ músculo\end{tabular}                                                                   & \multicolumn{3}{c}{bíceps braquial direito} \\ [1pt] \hline
\begin{tabular}[c]{@{}l@{}}Duração das\\  contrações\end{tabular}        & \multicolumn{3}{c}{$34$ $\mathrm{segundos}$}               \\ [1pt] \hline
\begin{tabular}[c]{@{}l@{}}Tempo médio \\ de descanso\end{tabular}        & \multicolumn{3}{c}{$5$ $\mathrm{minutos}$}               \\ [1pt] \hline
\begin{tabular}[c]{@{}l@{}}Tipo de \\ contração \end{tabular}                                                                 & isométrica      & alternada      & -        \\ [1pt] \hline
\begin{tabular}[c]{@{}l@{}}Ângulo da\\ articulação \end{tabular} & $70^{\circ}$             & $90^{\circ}$            & $120^{\circ}$     \\ [1pt] \hline
\begin{tabular}[c]{@{}l@{}}Massa do\\ halter \end{tabular}                                                               & 8 $\mathrm{kg}$            & 15 $\mathrm{kg}$          & -  \\[1pt] \hline 

\end{tabular}
\end{table}
O método de correlação cruzada estima o atraso entre dois sinais mediante a similaridade entre pontos de referência dos sinais. Sendo que qualquer ponto de referência, como picos, zeros e degraus dos sinais podem ser usados para alinhar os sinais e estimar o atraso. Desta forma, é praticável somente analisar o envelope das amplitudes dos sinais, considerando os limites positivos e negativos dos sinais EMGs.  

Os métodos de modulação em fluxo de bits e correlação cruzada podem ser aplicados para estimar a distância entre os pontos de referência. A conversão dos sinais em valores discretos utilizando limiares da amplitude é uma alternativa para condicionar o sinal, obtendo distintas representações dos sinais. No entanto, as representações ainda preservam informações fundamentais para a estimação do atraso utilizando correlação cruzada e posteriormente o cálculo da velocidade de condução.    

\subsection{Estimação da Velocidade de Condução das Fibras Musculares ($\mathrm{VCFM}$)}

A $\mathrm{VCFM}$ é uma medida da velocidade de deslocamento dos potenciais de ação das unidades motoras (MUAPs) no tecido muscular. A Equação \ref{eq_vcfm}
representa o cálculo do $\mathrm{VCFM}$:
    \begin{equation}
    \label{eq_vcfm}
    \mathrm{VCFM}= \left(\frac{\mathrm{D}}{\Delta t}\right)\mathrm{F},
\end{equation}
\noindent em que $\mathrm{D}$ é a distância entre os eletrodos, $\Delta t$ é o atraso entre os sinais derivados dos canais (eletrodos) espaçados e $\mathrm{F}$ é a frequência de amostragem dos dados coletados. A $\mathrm{VCFM}$ é dado em metros por segundo ($\mathrm{m/s}$) e o seu cálculo depende da estimação do atraso $\Delta t$ em tempo discreto das amostras derivadas de dois sinais EMGs.

O sinal EMGs representado em fluxo de bits é segmentado em janelas de tempo variável de $1$ segundo, fornecendo apenas uma estimativa de $\mathrm{VCFM}$ em cada segmento. A estimativa da $\mathrm{VCFM}$ em todas as amostras é obtida pela movimentação da janela a cada $2200$ amostras. A cada janela ($\mathrm{L}$) é calculada a função correlação cruzada, em que é o índice do elemento do vetor que possui correlação máxima ($C_{max}$) é usado para estimar o atraso $\Delta t$, representado pela seguinte equação:   
\begin{equation}
    \Delta t = C_{\mathrm{max}} - \mathrm{L},
\end{equation}
\noindent em que $C_{\mathrm{max}}$ é a posição da amostra em que a função de correlação é máxima e $\mathrm{L}$ é o tamanho da janela. Desta forma, os valores de atraso $\Delta t$ e a distância conhecida dos eletrodos possibilita a estimação da velocidade de condução das fibras musculares pela Equação \ref{eq_vcfm}.

\section{Resultados}

\begin{figure}[ht]
	\begin{minipage}[c][0.5\width]{
	   0.5\textwidth}
	   \centering
	   \includegraphics[width=1\textwidth]{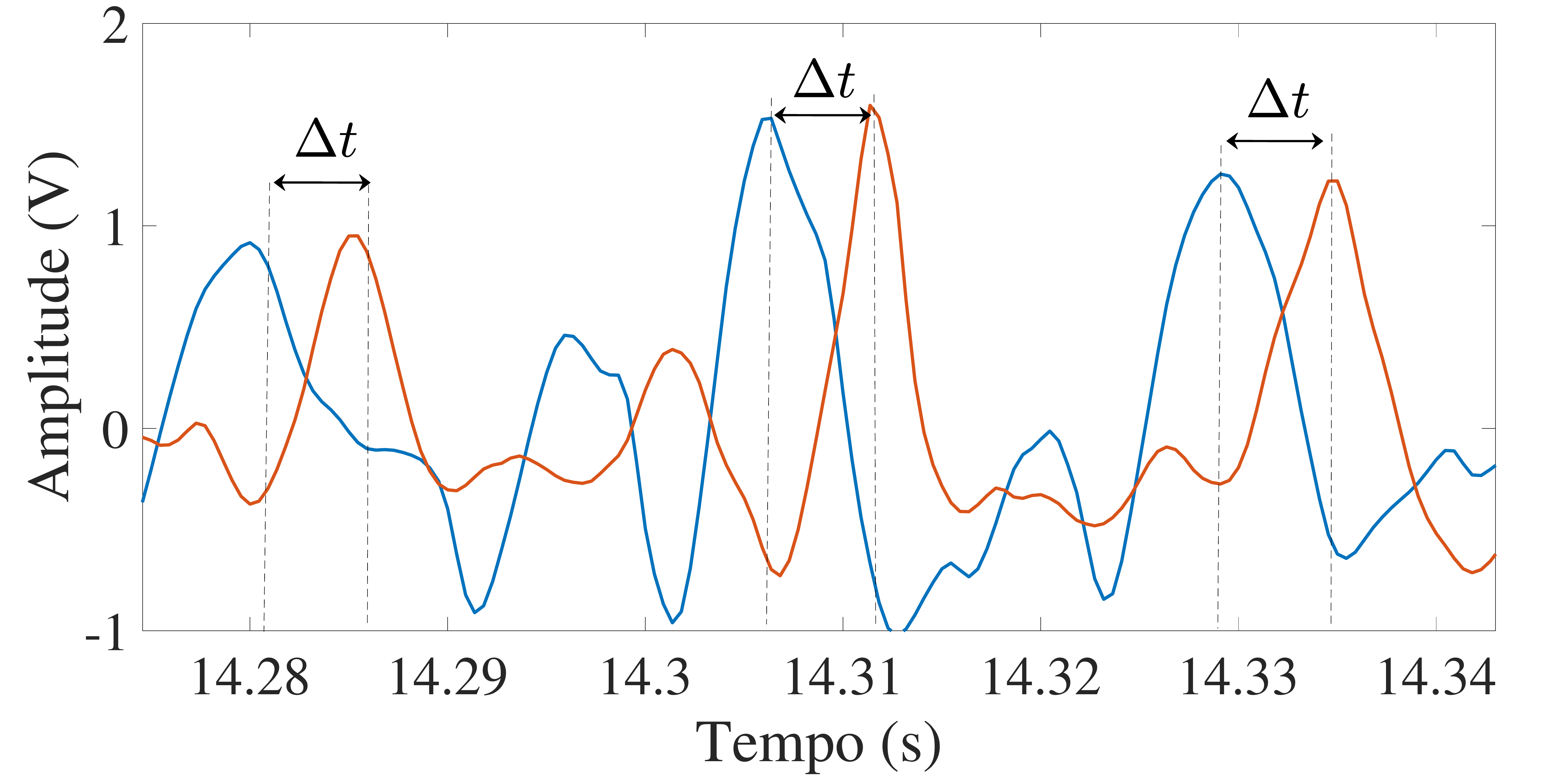}
	\end{minipage}
\caption{ Comparação dos sinais EMGs derivados dos canais A e B, mostrando a similaridade dos sinais e a representação do atraso $\Delta t$.}
\label{fig:comp_A_B}
\end{figure}

Para análise dos resultados foram utilizados os dados de EMGs relacionados ao experimento realizado com os seguintes parâmetros: contração isométrica, ângulo de articulação de $90^{\circ}$ e massa do halter de $15$ $\mathrm{kg}$ mostrados na Tabela \ref{tab:param_exp}. Primeiramente é verificado se os dados coletados exibem similaridade suficiente para calcular o atraso. Mediante o algoritmo de correlação cruzada a função resultante exibe um alto grau de similaridade. A Figura \ref{fig:comp_A_B} mostra graficamente os sinais de EMGs coletados dos dois canais, em que os sinais exibem alta similaridade e um atraso $\Delta t$.

\begin{figure}[ht!]
	\begin{minipage}[c][0.5\width]{
	   0.5\textwidth}
	   \centering
	   \includegraphics[width=1\textwidth]{mod_A_bit_G_1}
	\end{minipage}
	\begin{minipage}[c][0.5\width]{
	   0.5\textwidth}
	   \centering
\includegraphics[width=1\textwidth]{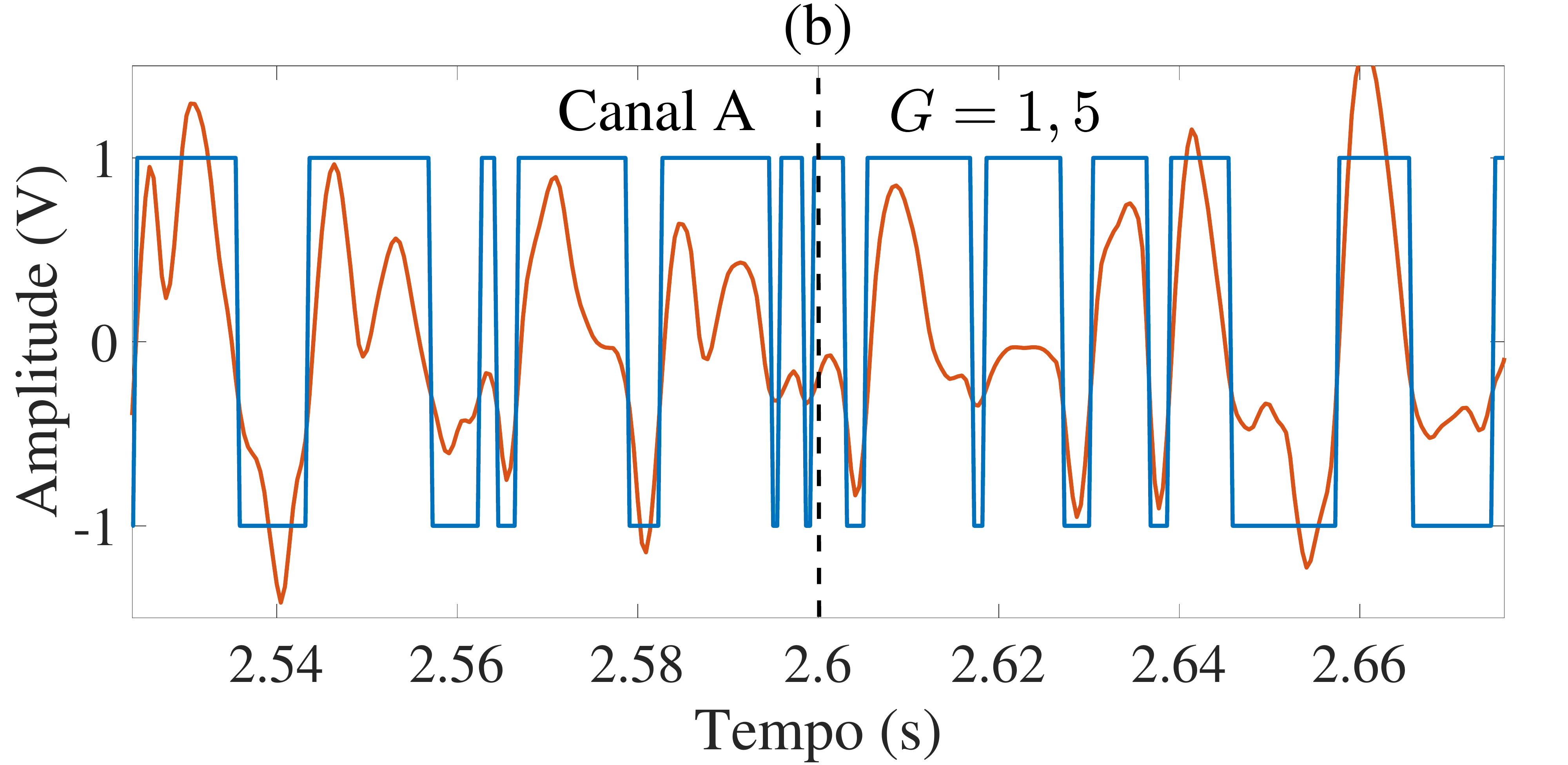}
	\end{minipage}
\caption{Representações da modulação em fluxo de bits do canal A com ganho de modulação variável, (a) representa o sinal modulado com $G=1$ e (b) o sinal modulado com $G=1,5$, mostrando duas representações do sinal EMGs.}
\label{fig:comp_ganho}
\end{figure}

\begin{table*}[ht!]
\centering
\setlength{\tabcolsep}{15pt} 
\renewcommand{\arraystretch}{1.1}
\caption{Valores estimados de \textrm{VCFM} comparando o método proposto com os valores obtidos na literatura, mostrando os intervalos, média e desvio padrão ($\mathrm{SD}$) derivados da análise de contrações isométricas do bíceps braquial direito. É observado que a estimativa exibe valores de VCFM compreendidos nos intervalos fornecidos pela literatura e valores médios similares.}
\label{tab:comp_est_vel}
\begin{tabular}{l c c c}
\hline 
 \textbf{Referência/Método} & \multicolumn{1}{c}{\begin{tabular}[c]{@{}c@{}}MFCV\\Intervalo\\(m/s)\end{tabular}} & \multicolumn{1}{c}{\begin{tabular}[c]{@{}c@{}}MFCV \\ média $\pm \ \mathrm{SD}$  \\ (m/s)  \end{tabular}} & \multicolumn{1}{c}{\begin{tabular}[c]{@{}c@{}}Níveis de contração isométrica: \\ bíceps braquial direito \end{tabular}} \\ [1pt] \hline 
\multicolumn{1}{l}{\begin{tabular}[l]{@{}l@{}}\textbf{Proposto} \\ Correlação cruzada com \\ fluxo de bits  \textit{software} \end{tabular}} &  $[4,1 - 5,5]$  & $4,5 \pm 0,6$  & \multicolumn{1}{l}{\begin{tabular}[l]{@{}l@{}} $90^{\circ}$ em relação a articulação\\ do cotovelo.\end{tabular}} \\[1pt] \hline  \multicolumn{1}{l}{\begin{tabular}[l]{@{}l@{}}\textbf{\cite{Koutsos2016}} \\ Correlação cruzada com \\ fluxo de bits  ASICs\end{tabular}} &  $[3,5 - 5,6]$  & $4,5 \pm 0,8 $  & \multicolumn{1}{l}{\begin{tabular}[l]{@{}l@{}} $70\%$ da máxima contração \\ voluntária. \end{tabular}} \\[1pt] \hline
 \multicolumn{1}{l}{\begin{tabular}[l]{@{}l@{}} \textbf{\cite{Marco2017}} \\ Correlação cruzada\end{tabular}} &  $[3,0 - 4,2]$  & $4,1 \pm 0,2 $ & \multicolumn{1}{l}{\begin{tabular}[l]{@{}l@{}} $60\%$ da máxima contração \\ voluntária. \end{tabular}} \\[1pt] \hline
 \multicolumn{1}{l}{\begin{tabular}[l]{@{}l@{}} \textbf{\cite{Xu2017}} \\ Detecção de zero \\ \textit{Delay-locked loop} \end{tabular}} &  $[2,6 - 4,3]$  & $3,3 \pm 0,3 $  & \multicolumn{1}{l}{\begin{tabular}[l]{@{}l@{}} $80\%$ da máxima contração \\ voluntária. \end{tabular}} \\[1pt] \hline
 \multicolumn{1}{l}{\begin{tabular}[l]{@{}l@{}} \textbf{\cite{Ye2015}} \\ Correlação cruzada\end{tabular}} &  $[2,5 - 4,9 ]$  & $3,3 \pm 0,7 $  & \multicolumn{1}{l}{\begin{tabular}[l]{@{}l@{}} $120^{\circ}$ articulação do cotovelo\\ $60\%$ da máxima contração. \end{tabular}} \\[1pt] \hline
 \multicolumn{1}{l}{\begin{tabular}[l]{@{}l@{}} \textbf{\cite{Farina2004}} \\ Correlação Cruzada\end{tabular}} &  $[3,1 - 4.9]$  & -  & \multicolumn{1}{l}{\begin{tabular}[l]{@{}l@{}}  $50\%$ da máxima contração \\ voluntária. \end{tabular}} \\[1pt] \hline
\end{tabular}
\end{table*}

Verificada a existência do atraso entre os sinais é realizado a aplicação da modulação em fluxo de bits variável. Os sinais são processados com ganho de modulação ($G$) entre $1$ e $1,5$, resultando em diferentes representações dos sinais. A Figura \ref{fig:comp_ganho} mostra duas representações de modulação do canal A com diferentes ganhos de modulação. Como mostrado nos gráficos o sistema se torna mais receptivo, em que os picos extraídos se tornam mais amplos e os picos menores são captados.

\begin{figure}[ht]
	\begin{minipage}[c][0.5\width]{
	   0.5\textwidth}
	   \centering
	   \includegraphics[width=1\textwidth]{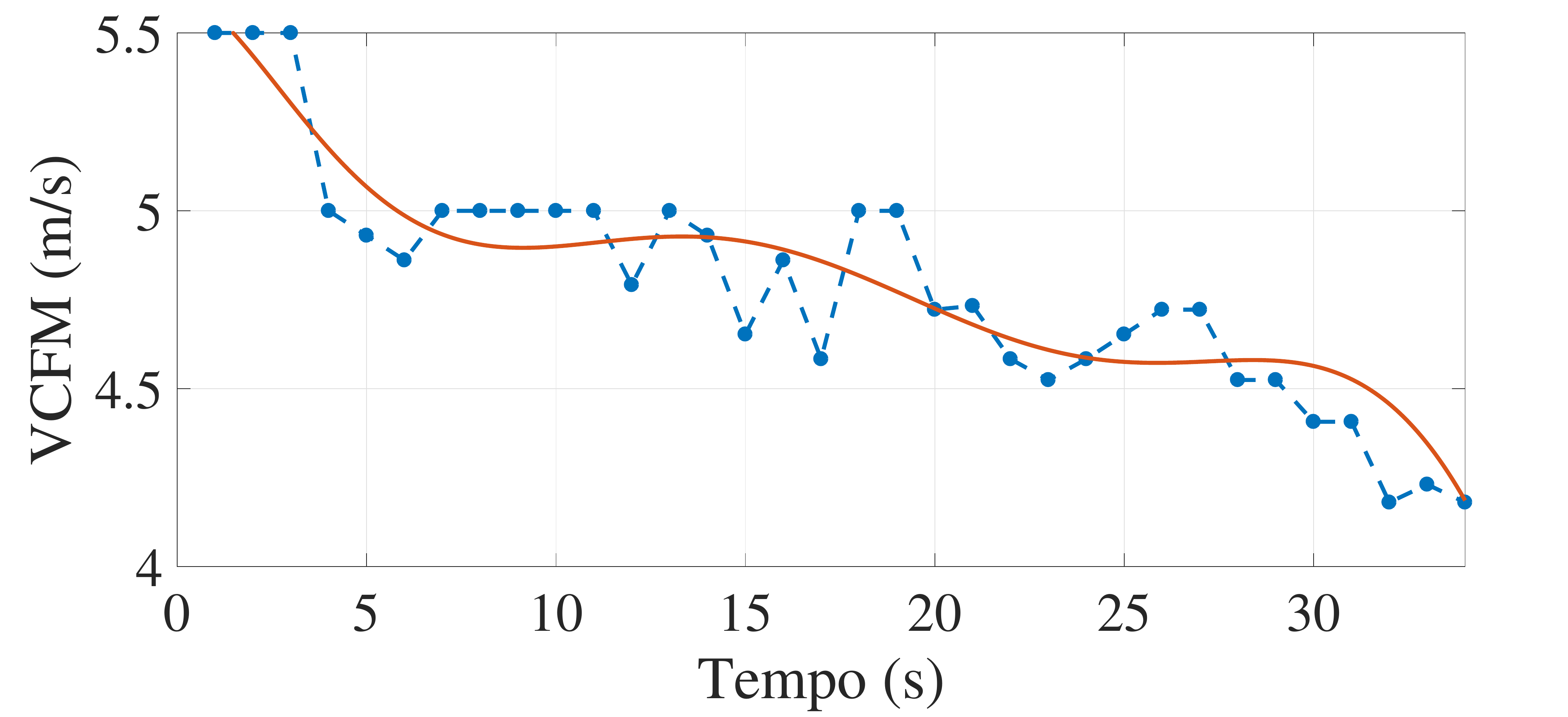}
	\end{minipage}
\caption{Representação das médias calculadas a cada segundo pelas distintas representações da $\mathrm{VCFM}$, em que foi utilizado a modulação de fluxo de bits variável. É destacado a redução progressiva da VCFM, correlacionando a tendência com o estado de fadiga muscular.}
\label{fig:vcfm_mod}
\end{figure}

A segunda etapa é estimar o atraso entre os sinais pelo método de correlação cruzada, utilizando as diferentes representações de modulação derivadas dos ganhos. Desta forma, foi calculado a $\mathrm{VCFM}$ para cada atraso estimado, obtendo um conjunto da $\mathrm{VCFM}$ para cada instante de tempo. Para garantir uma melhor estimação da velocidade de condução foi calculada a média e desvio padrão dos conjuntos de dados. A Figura \ref{fig:vcfm_mod} apresenta a estimação da $\mathrm{VCFM}$ durante $34$ segundos, mostrando que a $\mathrm{VCFM}$ reduz progressivamente para valores mais baixos, correlacionando a tendência observada com o estado de fadiga muscular. Os sinais de EMGs são adquiridos no processo de desgaste do músculo braquial direito durante contrações isométricas, em que sinais com amplitudes menores que $0,2$ $\mathrm{V}$ foram excluídos para indicar os sinais reais de propagação e não apenas ruídos. Na estimativa VCMF também foram examinados valores menores que $6$ $\mathrm{m/s}$, pois valores acima não são fisiologicamente razoáveis para $\mathrm{VCFM}$.

Para validar os valores obtidos de $\mathrm{VCFM}$, alguns critérios são comparados com dados da literatura, como o intervalo dos valores encontrados, média e desvio padrão. A Tabela \ref{tab:comp_est_vel} apresenta os valores obtidos pelo método proposto e valores fornecidos pela literatura, em que os métodos analisam a contração isométrica do bíceps braquial direito. A diferença entre os experimentos investigados consiste nos níveis de contração e ângulo da articulação. 

Os valores obtidos utilizando a combinação de correlação cruzada e modulação em fluxo de bits, consistem em um intervalo de $\mathrm{VCFM}$ entre $4,1$ e $5,5$ $\mathrm{m/s}$ para contrações isométricas. O intervalo estimado está incluso nos intervalos fornecidos pela literatura, porém incorporando incertezas relativas a sensibilidade de ruídos. A Tabela \ref{tab:comp_est_vel} mostra que o limite inferior do intervalo proposto possui valor menor em comparação com a literatura, indicando uma maior sensibilidade ao ruído ao decorrer que o atraso entre os sinais aumenta. Desta forma, o decréscimo da velocidade foi melhor avaliado considerando uma faixa possível de graus de similaridade dos sinais.

\section{Conclusão}

O presente trabalho apresenta um método não invasivo para estimativa da velocidade de condução das fibras musculares, que consiste no uso de técnicas como correlação cruzada e modulação em fluxo de bits. A principal contribuição do trabalho é o uso da modulação variável para estimação da velocidade de condução, sendo possível distinguir estatisticamente a diminuição da $\mathrm{VCFM}$ estimada em diferentes representações de modulação dos sinais EMGs. O método foi avaliado mediante estudo experimental e forneceu estimativas da $\mathrm{VCFM}$ similares às obtidas pela literatura \citep{Koutsos2016}, fornecendo valores mais conservadores ao considerar a média de um conjunto de velocidades a cada segundo. 

Para a estimação foram necessárias janelas extensas de amostras, produzindo apenas uma estimativa de velocidade a cada janela. No entanto, considerando um monitoramento contínuo a estimativa deve ser alcançada mediante a movimentação das janelas a cada amostra. Assim, o método proposto não pode ser usado para uma avaliação em tempo real da fadiga muscular, como também é necessário uma avaliação em diferentes tipos de músculos, sendo que o método proposto tem como objetivo a avaliação da fadiga para o bíceps braquial. Dentro das limitações, as técnicas apresentadas neste trabalho representam uma ferramenta não invasiva que abre perspectivas interessantes em estudos do monitoramento dos músculos, fornecendo meios para interpretar manifestações mioelétricas.

Em trabalhos futuros, pretende-se construir um protótipo embarcado vestível para o monitoramento e análise contínua da fadiga muscular. O protótipo pode ser usado para avaliação de ergonomia, esportes e condições de trabalhadores submetidos a exercícios repetitivos e sustentados.

\section*{Agradecimentos}
Agradecemos à CAPES, CNPq, INERGE, FAPEMIG e à Universidade Federal de São João del-Rei pelo apoio. 

\bibliography{ifacconf}           

\end{document}